\documentclass{elsart}
\usepackage{epsfig}

\begin{document}

\begin{frontmatter}
\title{Extended Brown-Rho Scaling Law with QCD Sum Rules}
\author{L. A. Barreiro}
\address{Instituto de Geoci\^encias e Ci\^encias Exatas,\\
Departamento de F\'{\i}sica - UNESP \\
Cx.P. 178, CEP 13500-970, Rio Claro, SP, Brazil. \\
barreiro@rc.unesp.br}

\begin{abstract}
The scalar and vector self-energies obtained through QCD sum rules
are introduced in Quantum Hadrodynamics (QHD) equations. The
results indicate that the ratios of coupling constants to
respective meson mass have a very small dependence on density.
Then, an extended Brown-Rho scaling law is conjectured.
\end{abstract}
\begin{keyword}
Nuclear Matter, QCD sum rules and Self-energies%
\PACS{21.65.+f, 24.85.+p}%
\end{keyword}
\end{frontmatter}

In recent years the density dependence of meson masses and
meson-baryon coupling constants have received increasing interest
\cite {BRscal,BrokTok,SoMBR,LiLeeBR}. Medium modification of
hadron properties can be seen as a manifestation of quark
substructure. One of the principal reasons to study that topic is
the success of Quantum Hadrodynamics (QHD) \cite{SeW,SeW2} in the
relativistic description of nuclear phenomena. QHD has large
success in the explanation of properties of infinite nuclear
matter as well as finite nuclei. Several calculations of nuclear
structure using QHD and its extensions were made obtaining a good
agreement with experimental data \cite{SeW2,LSR}.

QHD describes the N-N (nucleon-nucleon) interaction through mesons exchange (%
$\pi $, $\sigma $, $\omega $, $\rho $, etc.). However, for the
purposes of this work, only the simplest QHD model, namely QHD-I,
which includes nucleons ($\psi $) coupled with sigma ($\sigma $)
and omega\ ($\omega $) mesons, will be considered. In spite of
the pion to be the principal component of the N-N interaction, it
does not play any role in the model because nuclear matter is an
isotropic system with parity conservation. Analyzing the model,
it is possible to verify that the real part of the scalar term
($\sigma $-meson) is typically of the order of several hundred MeV
attractive while the real part of\ the time component of the vector term ($%
\omega $-meson) is typically of the order of several hundred MeV
repulsive. However, the energies involved in problems of nuclear
structure are only of a few tens of MeV. An energy of that order
is obtained in QHD models due to a large cancellation between the
scalar and vector pieces. In fact, scalar mesons produce a long
range attractive potential which saturate due to the presence of
a repulsive short range potential produced by vector mesons. This
process of saturation can be controlled through an appropriate
choice of coupling constants. However, there are strong
indications that such constants, as well as meson masses, change
with the density \cite {BRscal,BrokTok}. So, the success of QHD,
which uses fixed values for these amounts, must be explained.

Recent results of the effective field theory (EFT) \cite
{SeW2,Wein1,Wein2,kaplan,Furnst1,Furnst2} have verified that QHD
models are consistent with symmetries of Quantum Chromodynamics
(QCD). Motivated by that, it will be very interesting to make a
connection between QHD and QCD with the intention of studying
coupling constants and meson masses in the medium. The natural
way to make that connection take into account the nucleon
self-energy. That idea was applied by Brockmann and Toki
\cite{BrokTok} to connect Relativistic Bruckner-Hartree-Fock
Theory and Relativistic Hartree Approach. However, as it is known,
the energies involved in nuclear matter problems are very lower
than the energy scale of perturbative QCD. Thus, it is necessary
to use a nonperturbative procedure. The indicated is the
well-known method of QCD sum rules, that was introduced by
Shifman, Vainshtein, and Zakharov in the late 1970's \cite{SVZ}.
The QCD sum rule approach was successfully used in a lot of
different problems, since hadron properties to particle decay.
For the purposes of this work, the most interesting use of QCD
sum rules is the possibility of describing properties of a
nucleon in the medium. Indeed, QCD sum rules succeed to obtain
the nucleon self-energy in terms of quarks and gluons degree of
freedom \cite{CFG1,CFG2,CFG3,CFG4,CFGX}. On the other hand,
self-energies obtained in QHD models are dependent on coupling
constants and meson masses. The objective of this work is to
study these hadronic parameters of a more fundamental point of
view.

To study meson-nucleon coupling constants and meson masses in
nuclear matter, it is first necessary to solve the nuclear ground
state within Relativistic Hartree Approximation. The energy
density in the QHD-I model can be written as \cite{SeW}
\begin{equation}
\varepsilon =\frac{1}{2}\frac{m_{s}^{2}}{g_{s}^{2}}\Sigma _{(s)}^{2}+\frac{1%
}{2}\frac{m_{v}^{2}}{g_{v}^{2}}\Sigma _{(v)}^{2}+\frac{\gamma }{(2\pi )^{3}}%
\int_{0}^{k_{F}}\mbox{d}^{3}k\sqrt{|{\bf k|}^{2}+M^{\ast 2}}\;,
\label{e-mft}
\end{equation}

where $M^*$ is the effective nucleon mass in matter, $k_{F}$ is
the Fermi momentum, $\gamma $ is the spin-isospin degenerescence,
$g_{s}$ and $g_{v}$ are, respectively, the $\sigma$-N and the
$\omega$-N coupling constants and $m_s$ and $m_v$ are,
respectively, the $\sigma$- and $\omega$-meson masses. The
self-energies $\Sigma _{(s)}$ and $\Sigma _{(v)}$ can be obtained
through ``tadpole'' Feynman diagrams resulting in
\begin{equation}
\Sigma _{(s)}=M^{\ast }-M=-\frac{g_{s}^{2}}{m_{s}^{2}}\rho _{s}
\label{sigma_s}
\end{equation}
and
\begin{equation}
\Sigma _{(v)}^{\mu }=\delta ^{\mu 0}\frac{g_{v}^{2}}{m_{v}^{2}}\rho _{B}\;,
\label{sigma_v}
\end{equation}
Here $M$ is the free nucleon mass with value $M\approx939.$ MeV.
The baryon and the scalar densities are, respectively, given by:
\begin{equation}
\rho _{B}=\frac{\gamma }{6\pi ^{2}}k_{F}^{3}
\end{equation}
and
\begin{equation}
\rho _{s}=\frac{\gamma M^{\ast }}{4\pi ^{2}}\left[ k_{F}E_{F}^{\ast
}-M^{\ast 2}\ln \left( \frac{k_{F}+E_{F}^{\ast }}{M^{\ast }}\right) \right]
\end{equation}

The self-energies (\ref{sigma_s}) and (\ref{sigma_v}) can also be
obtained through QCD sum rules. The principal ingredient of this
method is the time-ordered correlation function, defined by
\begin{equation}
\Pi _{\alpha \beta }(q)\equiv i\int \mbox{d}^{4}x\mbox{
e}^{iq\cdot x}\langle 0|T\left[ \eta _{\alpha
}(x)\bar{\eta}_{\beta }(0)\right] |0\rangle \;, \label{correlator}
\end{equation}
where $|0\rangle $ is the physical nonperturbative vacuum state and $\eta
_{\alpha }(x)$ is an interpolating field with the quantum numbers of a
nucleon. In agreement with Ref. \cite{CFGX}, we have for the proton field
\begin{equation}
\eta (x)=\epsilon _{abc}\left[ t\left( u_{a}^{T}C\gamma _{5}d_{b}\right)
u_{c}+\left( u_{a}^{T}Cd_{b}\right) \gamma _{5}u_{c}\right] \;,
\end{equation}
where $u$ and $d$ are the up and down quark fields, $a,$ $b,$ $c$ are color
indices, $T$ means transpose, $C$ is the charge-conjugation matrix and $t$
is an arbitrary real parameter. The OPE for the correlation function (\ref
{correlator}) can be generated through the following expansion for the quark
propagator\cite{RRY,KCY}
\begin{eqnarray}
S_{ab}(x) &=&\langle 0|T\left[ q_{a}(x)\bar{q}_{b}(0)\right] |0\rangle
\nonumber \\
&=&i\frac{\delta _{ab}}{2\pi ^{2}}\frac{\not{x}}{x^{4}}-\frac{\delta _{ab}}{%
12}\langle \bar{q}q\rangle _{vac}+\cdots \;,
\end{eqnarray}
where $\langle \bar{q}q\rangle _{vac}$ is the free quark condensate, which
can be determined from the Gell-Mann-Oakes-Renner relation,
\begin{equation}
2m_{q}\langle \bar{q}q\rangle _{vac}=-m_{\pi }^{2}f_{\pi }^{2}\left( 1+{ %
O}(m_{\pi }^{2})\right) \;.  \label{vacondens}
\end{equation}
Here, $m_{\pi }=138$ MeV is the pion mass, $f_{\pi }=93$ MeV is the pion
decay constant and $m_{q}=(m_{u}+m_{d})/2$ is the average quark mass.

The same correlation function, Eq. (\ref{correlator}), can be described
through a phenomenological anzats inspired in the nucleon Green function.
With the fundamental state of the nuclear matter as vacuum and the
interacting propagator in the phenomenological side, we obtain the following
sum rules \cite{CFGX}
\begin{eqnarray}
M_{N}^{\ast } &=&-\left( \frac{7-2t-5t^{2}}{5+2t+5t^{2}}\right) \frac{%
M_{B}^{2}\langle \bar{q}q\rangle _{\rho _{B}}}{\frac{1}{16\pi ^{2}}%
M_{B}^{4}+\frac{2}{3}E_{q}\langle q^{\dagger }q\rangle _{\rho _{B}}%
}\;,  \label{ms} \\%
\Sigma _{(v)}^{QCD} &=&\frac{M_{B}^{2}\langle q^{\dagger
}q\rangle _{\rho _{B}}}{\frac{3}{64\pi
^{2}}M_{B}^{4}+\frac{1}{2}E_{q}\langle q^{\dagger }q\rangle
_{\rho _{B}}}\;.  \label{mv}
\end{eqnarray}
where $M_{B}^{2}$ represent the Borel mass with value near
$M_{B}^{2}\simeq 1 $GeV$^{2}$\cite{IOF}, $\langle q^{\dagger
}q\rangle _{\rho _{B}}={3\rho _{B}}/{2}$ represents the quark
density, $\langle \bar{q}q\rangle _{\rho _{B}}$\ is the in-medium
quark condensate \cite{CFG} and
$E_{q}=-\Sigma_{(v)}+\sqrt{\mbox{\boldmath $q$}^2+M_N^{*2}}$, with
$|\mbox{\boldmath $q$}|$=270 MeV ({\it i.e.}, approximately the
Fermi momentum). Here, it was not take into account the continuous
contribution. The expression (\ref{ms}), with $t=-1$, is a
generalization of the Ioffe's formula \cite{IOF} to finite
density. The explicit dependence on the choice of the parameter
$t$, that means, on the choice of the interpolating field, limits
the significance of the Ioffe's formula to a qualitative role at
best. A more realistic value for $t$ can be obtained in order to
do minimal the contributions coming from continuum model and high
dimensional operators in OPE \cite{Lei}. This better value will
be considered later. Calculating the ratio of finite-density to
zero-density sum rules, we obtain
\begin{eqnarray}
\frac{M_{N}^{\ast }}{M_{N}} &=&\frac{\langle \bar{q}q\rangle _{\rho _{B}}}{%
\langle \bar{q}q\rangle _{vac}}
\left( \frac{1}{1+{16 \pi ^{2}}\frac{E_{q}}{M_{B}^{4}}%
\rho _{N}}\right),  \label{self_qcd_s1} \\
\frac{\Sigma _{(v)}^{QCD}}{M_{N}} &=&
\frac{5+2t+5t^{2}}{7-2t-5t^{2}} \left(\frac{1}{1+16\pi ^{2}\frac{E_{q}}{%
M_{B}^{4}}{\rho _{N}}}\right) \frac{2}{\langle \bar{q}q\rangle
_{vac}}\rho _{N}\;. \label{self_qcd_v1}
\end{eqnarray}

Comparing expressions (\ref{sigma_s}) and (\ref{self_qcd_s1}), we
have for the scalar and vector sectors, respectively,
\begin{equation}
\frac{g_{s}^{2}}{m_{s}^{2}}=-\frac{M_{N}}{\rho _{s}}\left[
\frac{\langle \bar{q}q\rangle _{\rho _{B}}}{\langle
\bar{q}q\rangle _{vac}}\left( \frac{1}{1+{16 \pi ^{2}}\frac{E_{q}}{M_{B}^{4}}%
\rho _{B}}\right)-1\right]  \label{scalratio}
\end{equation}
and
\begin{equation}
\frac{g_{v}^{2}}{m_{v}^{2}}=-M_{N}\left[ \left( \frac{5+2t+5t^{2}}{%
7-2t-5t^{2}}\right) \frac{2}{\langle \bar{q}q\rangle
_{vac}}\frac{1}{1+16\pi ^{2}\frac{E_{q}}{M_{B}^{4}}\rho
_{B}}\right] . \label{vecratio}
\end{equation}

To analyze the expressions (\ref{self_qcd_s1}-\ref{vecratio}) is
necessary to know the in-medium quark condensate, which can be
written as \cite{CFG}
\begin{equation}
\langle \bar{q}q\rangle _{\rho _{B}}=\left( 1-\frac{\sigma _{N}\rho _{B}}{%
m_{\pi }^{2}f_{\pi }^{2}}+\cdots \right) \langle \bar{q}q\rangle _{vac}\;,
\label{quarkcondens}
\end{equation}
where $\langle \bar{q}q\rangle _{vac}$ is given by
Eq.(\ref{vacondens}) and the sigma term is estimated in Ref.
\cite{GLS} as being $\sigma _{N}\simeq 45\pm 10$ MeV. The
expression (\ref{quarkcondens}) is only been worth for low
densities and considering those conditions we can write in a rude
way the following approaches:  $M_{B}^{4}\gg E_{q}{\rho _{B}} $
and $\rho _{B}\sim \rho _{s}$. Therefore, the scalar sector is
written as
\begin{eqnarray}
\Sigma _{(s)}^{QCD} &=&-\frac{\sigma _{N}M_{N}}{m_{\pi }^{2}f_{\pi }^{2}}%
\rho _{B}\;,  \label{self_qcd_s2} \\
\frac{g_{s}^{2}}{m_{s}^{2}} &=& \frac{\sigma _{N}M_{N}}{m_{\pi
}^{2}f_{\pi }^{2}} \frac{\rho_B}{\rho_s} \approx \frac{\sigma
_{N}M_{N}}{m_{\pi }^{2}f_{\pi }^{2}} \;. \label{scalratioidepdens}
\end{eqnarray}
With the Ioffe's choice $(t=-1)$, we obtain for the vector sector,
\begin{eqnarray}
\Sigma _{(v)}^{QCD} &=&\frac{8m_{q}M_{N}}{m_{\pi }^{2}f_{\pi }^{2}}\rho
_{B}\;,  \label{self_qcd_v2} \\
\frac{g_{v}^{2}}{m_{v}^{2}} &=&\frac{8M_{N}m_{q}}{m_{\pi }^{2}f_{\pi }^{2}}%
\;.  \label{vecratioidepdens}
\end{eqnarray}
The principal conclusion obtained from equations (\ref
{scalratioidepdens}) and (\ref{vecratioidepdens}) is that the
ratios of coupling constants to respective meson mass are
constants for low densities. Some values for these ratios are
presented in table \ref{tab1}.
\begin{table}[h]
\begin{center}
\begin{tabular}{ccc}
\hline%
 & $g_{s}^{2}/m_{s}^{2}$ & $g_{v}^{2}/m_{v}^{2}$ \\ \hline
QHD-I & 3.029 10$^{-4}$ & 2.222 10$^{-4}$ \\
QCDSR & 2.793 10$^{-4}$ & 1.964 10$^{-4}$ \\ \hline%
\end{tabular}
\caption{ Ratios of coupling constants to respective meson mass.
In the first line are presented the values used in Ref.[5] and in
the second line the values obtained with equations
(\ref{scalratioidepdens}) and (\ref{vecratioidepdens}) using
Ioffe's interpolating field $(t=-1)$.
} %
\label{tab1}
\end{center}
\end{table}
The first line of that table presents the ratios used in Ref.
\cite{SeW}, which were adjusted so that QHD-I reproduces the
saturation properties of the nuclear matter. On the other hand,
when the self-energies (\ref {self_qcd_s2}) and
(\ref{self_qcd_v2}) and the ratios (\ref{scalratioidepdens}) and
(\ref{vecratioidepdens}) (low density limit) are applied in
Eq.(\ref{e-mft}), and we require the usual saturation condition
for nuclear matter, namely $\varepsilon/\rho_B - M = 15.75$ MeV
at normal nuclear matter density ($k_{F}=1.42$ fm$^{-1}$), we
obtain the values presented in the second line of table
\ref{tab1}. We can observe that the QCDSR results are similar to
the respective QHD-I values. However, this is not the most
important result. The in-medium behavior of $g_{s}^{2}/m_{s}^{2}$
and $g_{v}^{2}/m_{v}^{2}$, named scalar and vector ratios,
respectively, are the principal subject to be studied here.

\begin{figure}[h]
\centerline{\psfig{figure=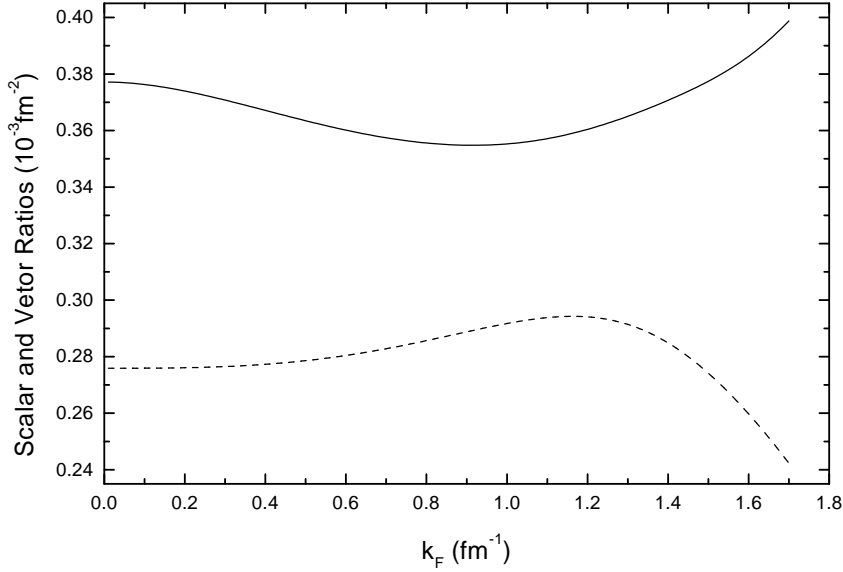,height=3.5in}}%
\caption{The ratios $g_s^2/m_s^2$ (solid line) and $g_v^2/m_v^2$
(dashed line) calculated with Eqs.(\ref{scalratio}) and
(\ref{vecratio}), respectively} \label{figure}
\end{figure}

Despite the fact of we have obtained the equilibrium properties
of nuclear matter, the effects despised in the low density
approximation must be included. For that, it is necessary to
abandon Eq.(\ref{quarkcondens}) for the condensate and to use the
familiar Nambu-Jona-Lasinio (NJL) model, which give us the
following expression \cite{CFG} for the quark condensate:
\begin{equation}
\langle \bar{q}q\rangle _{\rho _{B}}=-\frac{N_{c}M_{q}}{\pi ^{2}}%
\int_{k_{F_{(q)}}}^{\Lambda
_{NJL}}\mbox{d}p\frac{p^{2}}{\sqrt{M_{q}^{2}+p^{2}}} \label{NJL1}
\end{equation}
and
\begin{equation}
M_{q}=m_{q}-2G_{NJL}\langle \bar{q}q\rangle _{\rho _{B}}.  \label{NJL2}
\end{equation}
The Fermi momentum of quarks, $k_{F_{(q)}}$, is related to the
Fermi momentum of nucleons through the density: $\rho_q=3\rho_B$,
where $\rho_q$ is the quark density. Here $N_{c}$ is the color
number, $\Lambda _{NJL}$ is a cutoff, $G_{NJL}$ is a coupling
constant, $m_{q}$ is the current quark mass and $M_{q}$ is the
constituent quark mass, which is dynamically generated by a
partial restoration of chiral symmetry. With an appropriate
choice of these
parameters the NJL model gives us good results for the $\pi $-meson mass, $%
f_{\pi }$, constituent quark mass, condensate, etc. A good choice
is $\Lambda _{NJL}=900 $ MeV and $G_{NJL}=3.54/\Lambda
_{NJL}^{2}$, for which the saturation point of nuclear matter is
reproduced. The in-medium behavior of the relevant quantities
$g_{s}^2/m_{s}^2$ and $ g_{v}^2/m_{v}^2$, are presented on figure
\ref{figure}. The scalar [Eq. (\ref{scalratio})] and the vector
[Eq.(\ref{vecratio})] ratios are represented by the solid and the
dashed lines, respectively. In these calculation, a more
realistic nucleon interpolating field ($t=-1.1$) \cite{Lei} has
been used.

In summary, the expressions (\ref{self_qcd_s2}-\ref
{vecratioidepdens}) are valid just for the low density region.
However, some insight can be obtained from these first results:
the ratios of coupling constants to respective meson masses are
constants in the low density limit. Actually, the principal
conclusion obtained in this work can be summarized by figure
\ref{figure}, where the scalar and the vector ratios are
presented as functions of the density. We can observe that they
have a small dependence on density and the vector ratio decrease
when the density is enlarged. These results corroborate with the
observation made in Ref.\cite{LiLeeBR}, where the nucleon flow
probing higher density requires that $g_{v}/m_{v}$ be independent
of density at low densities and decrease slightly at high
densities. The results here obtained indicate that
$({g_{s}}/{m_{s}})_{vac}\approx ({g_{s}}/{m_{s}})_{\rho _{B}}$
and we can write
\begin{equation}
\frac{(m_{s})_{\rho _{B}}}{(m_{s})_{vac}}\approx \frac{(g_{s})_{\rho _{B}}}{%
(g_{s})_{vac}} ,  \label{result1}
\end{equation}
The same result is valid for the vector ratio. In other words,
coupling constants must change with the same rate of the
respective meson mass, in order to keep the ratios unaffected.
Therefore, the fact of the ratios remain constants can give us an
explanation of how QHD models, that use parameters independent of
density, have success to explain the bulk properties of nuclear
matter. The same is not true for finite nuclei, where the scalar
and vector ratios do not appear explicitly on equations. Thus, the
dependence on density of coupling constants and meson masses must
be important in these cases.

Equation (\ref{result1}) and its equivalent for the vector
sector, permit one to extend the Brown-Rho (BR) scaling law
\cite{BRscal} to include coupling constants in the following way:
\begin{equation}
\frac{M_{N}^{\ast }}{M_{N}}\approx \frac{f_{\pi }^{\ast }}{f_{\pi }}\approx
\frac{m_{s}^{\ast }}{m_{s}}\approx \frac{m_{v}^{\ast }}{m_{v}}\approx \frac{%
g_{s}^{\ast }}{g_{s}}\approx \frac{g_{v}^{\ast }}{g_{v}},
\label{BRscal}
\end{equation}
where the asterisks denote in-medium quantities.

Finally, it is necessary to say that these results are valid at
low densities. For nuclear matter in extreme conditions, the
ratios should change \cite{LiLeeBR} and QHD-I must present bad
results. Besides, in these calculations the contributions of
gluon condensates and other high terms were not included in the
OPE. However, we know that the contributions of these terms for
the sum rule are very small. The use of an extended version of
the sum rules (\ref{ms}) and (\ref{mv}) as well as the importance
of scaling for finite nuclei, are left for future works.

\noindent {\bf Acknowledgments:} I would like to thank FAPESP
(Funda\c{c}\~{a}o de Amparo \`{a} Pesquisa do Estado de S\~{a}o
Paulo) for the financial support.

\end{document}